\documentclass[10pt,aps,prl,twocolumn,superscriptaddress]{revtex4-2}

\usepackage{amsmath}
\usepackage{amsthm}
\usepackage{amssymb}
\usepackage{times}
\usepackage{braket}
\usepackage{ulem}

\usepackage{multirow}
\usepackage{array}
\usepackage{makecell}
\usepackage{diagbox}
\usepackage{booktabs}
\usepackage{nomencl}
\usepackage{siunitx}
\usepackage{tikz}
\usepackage{placeins}
\usepackage[utf8]{inputenc}
\usepackage[dvipsnames]{xcolor}
\usepackage{hyperref}
\hypersetup{
    colorlinks=true,
    linkcolor=Maroon,
    citecolor=Maroon,
    urlcolor=Maroon,
}

\usepackage{babel}

\makeatletter
\adddialect\l@en\l@english
\makeatother

\usepackage{graphicx}
\usepackage{svg}
\graphicspath{{fig/}}

\usepackage{algorithm, algpseudocode} 
\usepackage{mathrsfs} 
\usepackage{lipsum}

\usepackage{adjustbox}

\begin{document}

\title{Epitaxial single T centres in silicon-on-insulator}

\author{Christian H. Christiansen}
\altaffiliation{Equally contributing author}
\affiliation{NNF Quantum Computing Programme, Niels Bohr Institute, University of Copenhagen, Blegdamsvej 17, 2100 Copenhagen, Denmark.}

\author{Kasper H. Nielsen}
\altaffiliation{Equally contributing author}
\affiliation{NNF Quantum Computing Programme, Niels Bohr Institute, University of Copenhagen, Blegdamsvej 17, 2100 Copenhagen, Denmark.}

\author{Alisha Nanwani}
\altaffiliation{Equally contributing author}
\affiliation{Quantum Foundry Copenhagen, Copenhagen, Denmark.}

\author{Sebastiano Guaraldo}
\altaffiliation{Equally contributing author}
\affiliation{NNF Quantum Computing Programme, Niels Bohr Institute, University of Copenhagen, Blegdamsvej 17, 2100 Copenhagen, Denmark.}

\author{E. Laurits Piehorsch}
\affiliation{NNF Quantum Computing Programme, Niels Bohr Institute, University of Copenhagen, Blegdamsvej 17, 2100 Copenhagen, Denmark.}

\author{Arnulf J. Snedker-Nielsen}
\affiliation{NNF Quantum Computing Programme, Niels Bohr Institute, University of Copenhagen, Blegdamsvej 17, 2100 Copenhagen, Denmark.}

\author{Magnus L. Madsen}
\affiliation{NNF Quantum Computing Programme, Niels Bohr Institute, University of Copenhagen, Blegdamsvej 17, 2100 Copenhagen, Denmark.}

\author{David R. Gongora}
\affiliation{University of Oslo, Centre for Materials Science and Nanotechnology, PO Box 1048 Blindern, 0316, Oslo, Norway.}

\author{Emanuele Brusaschi}
\affiliation{Dipartimento di Fisica, Università di Pavia, Via Agostino Bassi 6, 27100 Pavia, Italy}
\affiliation{NNF Quantum Computing Programme, Niels Bohr Institute, University of Copenhagen, Blegdamsvej 17, 2100 Copenhagen, Denmark.}

\author{Rodrigo A. Thomas}
\affiliation{Quantum Foundry Copenhagen, Copenhagen, Denmark.}

\author{Ian Farrer}
\affiliation{Quantum Foundry Copenhagen, Copenhagen, Denmark.}

\author{D. Hieu Nguyen}
\affiliation{Quantum Foundry Copenhagen, Copenhagen, Denmark.}

\author{Georgios Kountouris}
\affiliation{NNF Quantum Computing Programme, Niels Bohr Institute, University of Copenhagen, Blegdamsvej 17, 2100 Copenhagen, Denmark.}

\author{Beñat M. d. A. Jokisch}
\affiliation{NNF Quantum Computing Programme, Niels Bohr Institute, University of Copenhagen, Blegdamsvej 17, 2100 Copenhagen, Denmark.}

\author{Mohammad Khalifa}
\affiliation{NNF Quantum Computing Programme, Niels Bohr Institute, University of Copenhagen, Blegdamsvej 17, 2100 Copenhagen, Denmark.}

\author{Claudia Piccinini}
\affiliation{NNF Quantum Computing Programme, Niels Bohr Institute, University of Copenhagen, Blegdamsvej 17, 2100 Copenhagen, Denmark.}

\author{Mathias Ø. Augustesen}
\affiliation{NNF Quantum Computing Programme, Niels Bohr Institute, University of Copenhagen, Blegdamsvej 17, 2100 Copenhagen, Denmark.}

\author{Hugo Laurell}
\affiliation{NNF Quantum Computing Programme, Niels Bohr Institute, University of Copenhagen, Blegdamsvej 17, 2100 Copenhagen, Denmark.}
\affiliation{Department of Physics, Lund University, Box 118, 22100 Lund, Sweden.}

\author{Kokeb B. Benti}
\affiliation{NNF Quantum Computing Programme, Niels Bohr Institute, University of Copenhagen, Blegdamsvej 17, 2100 Copenhagen, Denmark.}

\author{Maria S. Gonzalez}
\affiliation{NNF Quantum Computing Programme, Niels Bohr Institute, University of Copenhagen, Blegdamsvej 17, 2100 Copenhagen, Denmark.}

\author{Amedeo Carbone}
\affiliation{NNF Quantum Computing Programme, Niels Bohr Institute, University of Copenhagen, Blegdamsvej 17, 2100 Copenhagen, Denmark.}

\author{Elvedin Memisevic}
\affiliation{NNF Quantum Computing Programme, Niels Bohr Institute, University of Copenhagen, Blegdamsvej 17, 2100 Copenhagen, Denmark.}

\author{Sangeeth Kallatt}
\affiliation{NNF Quantum Computing Programme, Niels Bohr Institute, University of Copenhagen, Blegdamsvej 17, 2100 Copenhagen, Denmark.}

\author{Mark K. Svendsen}
\affiliation{NNF Quantum Computing Programme, Niels Bohr Institute, University of Copenhagen, Blegdamsvej 17, 2100 Copenhagen, Denmark.}

\author{Marianne E. Bathen}
\affiliation{University of Oslo, Centre for Materials Science and Nanotechnology, PO Box 1048 Blindern, 0316, Oslo, Norway.}

\author{Lasse Vines}
\affiliation{University of Oslo, Centre for Materials Science and Nanotechnology, PO Box 1048 Blindern, 0316, Oslo, Norway.}

\author{Peter Granum}
\affiliation{NNF Quantum Computing Programme, Niels Bohr Institute, University of Copenhagen, Blegdamsvej 17, 2100 Copenhagen, Denmark.}

\author{Peter Krogstrup}
\email{krogstrup@nbi.ku.dk}
\affiliation{Quantum Foundry Copenhagen, Copenhagen, Denmark.}
\affiliation{NNF Quantum Computing Programme, Niels Bohr Institute, University of Copenhagen, Blegdamsvej 17, 2100 Copenhagen, Denmark.}

\author{Stefano Paesani}
\email{stefano.paesani@nbi.ku.dk}
\affiliation{NNF Quantum Computing Programme, Niels Bohr Institute, University of Copenhagen, Blegdamsvej 17, 2100 Copenhagen, Denmark.}

\date{
	\today
}

\begin{abstract}
Spin-photon interfaces based on silicon quantum emitters offer a scalable platform for quantum computing and networking.
However, achieving coherent photon emission remains a primary challenge due to stringent material quality requirements.
To overcome this, we utilise high-purity molecular-beam epitaxy (MBE) to epitaxially incorporate single T centres in silicon-on-insulator (SOI) wafers.
We demonstrate single T-centre emission coupled to a nanophotonic waveguide and observe significant suppression of homogeneous broadening, yielding optical linewidths as narrow as 30 MHz using natural silicon for crystal growth.
These results establish epitaxial T centres as a robust foundation for coherent spin-photon interfaces in silicon quantum photonics.
\end{abstract}

\maketitle


%
Colour centres in silicon are emerging as a promising platform for solid-state spin-photon interfaces~\cite{redjem2020single, bergeron2020silicon, durand2021, simmons2024}.
They show excellent spin properties, can operate at telecom wavelengths, and can be directly integrated on silicon-on-insulator material stacks widely used for industry-scale chip manufacturing~\cite{lee2023high, deabreu2023waveguide, islam2023cavity, johnston2024cavity, sandholzer2026single}.
In particular, the T centre in silicon possesses a coherent electron spin directly coupled to the emission of photons in the telecom O-band, as well as to long-coherence nuclear spins forming additional memory qubits embedded in the defect~\cite{higginbottom2022, photinc2024distributed, song2026entanglement}.  
%
%

%
An important challenge in the development of this platform is the reduction of noise due to material imperfections, which currently limit the coherence of the optical and spin transitions in T centres. 
%
%
For example, photonic devices at 4~K  show typical linewidths for T centres of a few GHz due to spectral wandering, and hundreds of MHz due to homogeneous broadening, orders of magnitude worse than the linewidths in bulk silicon~\cite{deabreu2023waveguide}. 
Recent studies have associated these broadenings with the reconfiguration of the charge environment due to laser-induced charge transfer to or from additional defects near the T centre~\cite{Bowness2025LaserInduced, Zhang2025LaserInduced}.
The properties of T centres in silicon are indeed critically dependent on their spatial localisation and the crystallinity of the surrounding silicon matrix. 
The standard formation process, carbon and hydrogen ion implantation (see Fig.~\ref{fig:fig1}a-b), inevitably introduces a variety of additional defects alongside the desired T centres~\cite{macquarrie2021generating, snedker2026colour}. 
%
%
However, the coexistence of radiative and non-radiative defect states (e.g., vacancies, interstitial clusters, or impurity complexes) degrades the optical coherence of T centres by introducing competing charge states and recombination pathways~\cite{Bowness2025LaserInduced, Zhang2025LaserInduced}. 
This undesired defect landscape can lead to spectral diffusion and instability of the single-photon emission, ultimately limiting the performance of T centres in quantum applications.
Furthermore, ion implantation offers limited control over the depth at which colour centres are generated, with a depth distribution typically spanning a $\sim100$ nm region~\cite{aberl2024all}.
Recently, $\delta$-doping of carbon atoms has been demonstrated during epitaxial silicon growth via a low-temperature molecular beam epitaxy (MBE) process for all-silicon stacks. 
This approach was used for the self-assembled formation of ensembles of G-like, W, and T centres~\cite{aberl2024all, salomon2025telecom, marbock2026self}. 
It represents a promising alternative to ion implantation, offering nanometer-scale precision in defect positioning while avoiding the damage and disorder introduced by high-energy ion bombardment.
However, a key challenge lies in growing the $\delta$-doped layer and the surrounding silicon matrix to a purity matching float-zone silicon, within a material stack compatible with the manufacturing of nanophotonic devices. 
%
%

%
In this work, we develop a $\delta$-doping MBE process tailored to create single T centres in an SOI material stack, utilising a dedicated ultra-high-purity MBE system to address these challenges. 
We process the material to fabricate integrated waveguides and show the operation of individual epitaxial T centres in nanophotonic devices.
We characterise the epitaxial T centre devices and extract homogeneous linewidths as low as 30 MHz. This represents a ten-fold reduction in broadening compared to reference ion-implanted samples derived from the same SOI base substrate and evaluated under identical conditions. 

\section{Epitaxial formation of single T centres}
\label{sec:epigrowth}

\begin{figure*}
    \centering
\includegraphics[width=1.0\linewidth]{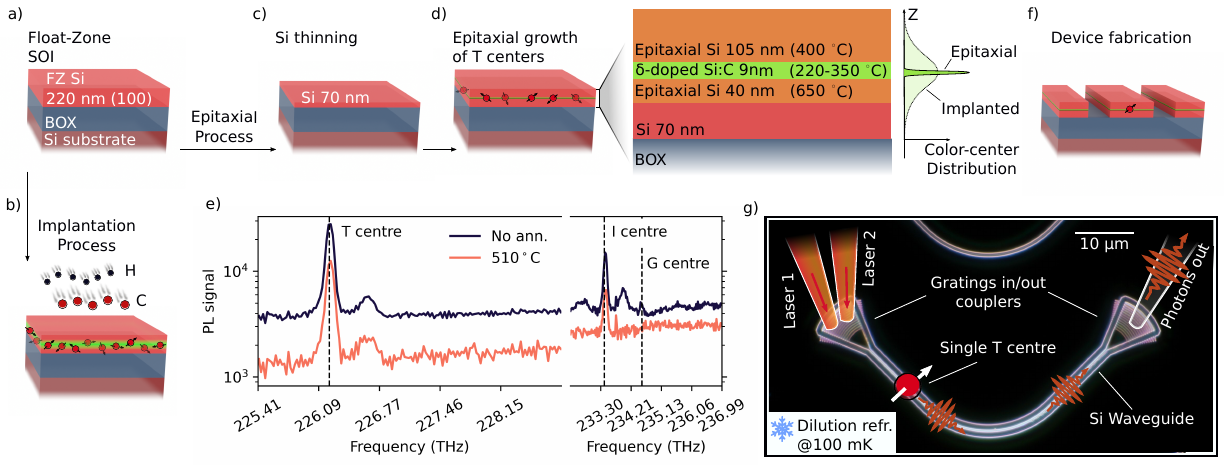}
    \caption{
    \textbf{Epitaxial generation of colour centres in SOI.}
    a) Cross-section of the initial SOI wafer with a top layer of 220~nm float-zone (FT) silicon.
    b) Standard implantation of T centres is performed through ion implantation of hydrogen (H) and carbon (C), causing damage to the silicon crystal, which can result in additional unwanted defects in the device.
    The process for growth-based generation of T centres in SOI starts with (c) thinning the FZ Si layer by back-etching, followed by (d) epitaxial growth of a stack of silicon layers.
    Inset of d): Schematic of the epitaxial process for growth-based generation of T centres. 
    Colour centres are incorporated through a low-temperature growth of a $\delta$-doped 9 nm Si:C layer, followed by capping with a silicon layer until the targeted total thickness of approximately $220$~nm is reached. 
    The depth of the generated T centres, whose distributions are depicted as green plots on the right side of d), is narrowly localised at the depth of the $\delta$-doped layer for the epitaxial method, in contrast to a distribution spread over $\gtrsim100$~nm for the implanted case~\cite{aberl2024all}.
    e) Above-band photoluminescence signal from T (left), I, and G (right) centres in the epitaxially grown material stack for an as-grown sample (black line), and for a sample that has been thermally annealed at $510~^\circ$C after growth (orange). 
    f) The wafers are finally processed to fabricate nanophotonic devices.
    g) Optical microscope image of a typical integrated device used to interface with T centres through integrated waveguides in the grown SOI stack. 
    %
    %
    Devices are operated in a dilution refrigerator at a typical temperature of 100~mK.
    } 
    \label{fig:fig1}
\end{figure*}

To integrate T centres with nanophotonic devices, it is key to generate them in a silicon-on-insulator material stack, where the thin top layer of silicon can be processed to form waveguides and other nanostructures. 
This has so far been achieved through standard implantation of commercial SOI wafers, where ion beam energies are tuned so implanted ions remain localised in the top thin-film silicon layer (see Fig.~\ref{fig:fig1}a-b)~\cite{macquarrie2021generating,snedker2026colour}.
Here, we develop a process for the creation of T centres epitaxially on the SOI stack, which is schematised in Fig.~\ref{fig:fig1}c-d.
%
%
We start by thinning the top 220~nm-thick silicon layer, nominally (001)-oriented, of a standard float-zone (FZ) $4^{\prime\prime}$ SOI wafer to a thickness of 70~nm using a back-etching process.
The wafer is then inserted into a dedicated high-purity MBE chamber for epitaxial growth of colour centres in silicon. 
In the MBE, we first grow an additional silicon buffer layer at a temperature of 650~$^\circ\text{C}$, sufficient to maintain good crystal quality, until the silicon reaches an additional thickness of approximately 40~nm.
The substrate temperature is then ramped down to a lower value in the range of  220-350~$^\circ\text{C}$ to facilitate carbon incorporation under kinetically limited conditions.
The $\delta$-doped layer is then added by depositing a 9~nm layer of carbon-doped silicon, namely Si:C, at the set temperature.
This thin Si:C layer is where a high concentration of carbon atoms is generated, which can recombine with diffusing hydrogen atoms naturally present in the material to form the self-assembled T centres in the $\delta$-doped region~\cite{aberl2024all}.
Finally, a silicon capping layer, with a thickness of 105~nm, is grown on top of the Si:C layer to complete the material stack. 
The capping process is carried out at a slightly higher growth temperature of approximately 400~$^\circ\text{C}$, to ensure good crystal quality while being sufficiently low to avoid the carbon atoms diffusing out of the $\delta$-doped Si:C layer.

\begin{figure*}
    \centering
    \includegraphics[width=1.0\linewidth]{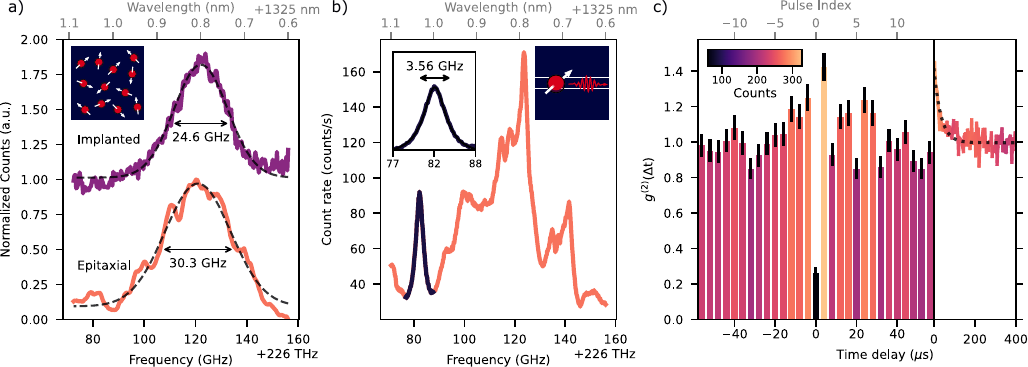}
    \caption{
    \textbf{Optical characterisation of epitaxial T centres.}
    a) Resonant photoluminescence spectra for ensembles of T centres in an epitaxially grown sample (orange), as well as from a standard implanted sample (purple). 
    An offset along the vertical axis has been added to the implanted spectrum to facilitate visualisation.  
    b) Resonant excitation spectrum collected in transmission from a single waveguide device in an epitaxially grown SOI sample.
    The emission of a single T centre can be spectrally resolved. The peak is highlighted in dark colour. 
    The rest of the spectrum, in orange, is from other T centres in the approximately 100~$\mu$m-long waveguide, spectrally separated from the single emitter being investigated. 
    c) Second-order autocorrelation measurement $g^{(2)}(\Delta t)$ obtained with resonant excitation of the single T centre.
    Normalising the side peaks for pulse indices associated with long delays yields an estimated autocorrelation of the zero-delay peak, $g^{(2)}(0)=0.26(3)$.
    The rightmost panel shows the $g^{(2)}(|\Delta t|)$ of the side peaks at large time delays. 
    An exponential decay of the signal with a decay time of 34~$\mu$s due to the spectral wandering of the emitter can be observed.
    Error bars represent 1$\sigma$ of the measured second-order autocorrelation values $g^{(2)}(\Delta t)$ and consider Poissonian photon statistics.
    }
    \label{fig:fig2}
\end{figure*}

%
%
In Figure~\ref{fig:fig1}e, we show photoluminescence (PL) signals obtained from the epitaxial SOI material, measured at a sample temperature of 4~K. 
Emission from T centres is visible, centred at the expected wavelength of $1325.8$~nm, already in the as-grown sample (black line).
The PL measurements also show the presence of other colour centres produced in the $\delta$-doped region, although with a smaller signal than the T centres.
These include G centres, a type of colour centre which has also shown promising properties as a quantum emitter in silicon (see right panel of Fig.\ref{fig:fig1}e)~\cite{beaufils2018optical, redjem2020single, komza2024indistinguishable},
and I centres, thought to be an oxygen-perturbed variant of the T centre~\cite{Gower_1997}.
%
%
We also report PL data from a sample that has been thermally annealed at 510$^\circ$C after growth (orange line), an annealing temperature value close to the optimal one for implanted material~\cite{macquarrie2021generating, snedker2026colour}.
Annealing was introduced to facilitate the recombination of carbon and hydrogen atoms to form T centres, as well as to thermally cure the material (see Appendix~\ref{sec:supp:anneal} for more details).
Note that, at these temperatures, the diffusion of carbon atoms in silicon is very limited, while hydrogen atoms can diffuse for tens of nanometers~\cite{snedker2026colour}.
As a result, the carbon atoms, and thus the generated T centres, remain localised within the $\delta$-doped layer, also in the presence of the additional annealing step. 
%
%
%
In the experiments reported in the rest of the paper, we use epitaxially grown samples annealed at a temperature of 510$^\circ$C, which provide a T-centre density suitable for addressing individual emitters in waveguides.


\section{Waveguide-integrated epitaxial T centres}
\label{sec:g2}

\begin{figure*}
    \centering
    \includegraphics[width=1.0\linewidth]{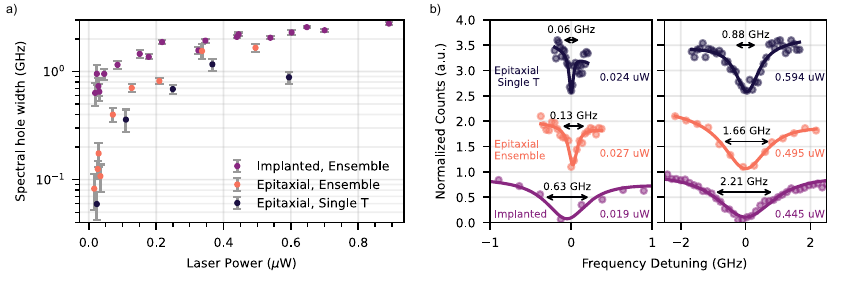}
    \caption{
    \textbf{Characterisation of the homogeneous broadening of the optical linewidth}.
    a) Instantaneous homogeneous optical linewidths for waveguide-integrated T centres measured through hole-burning.
    Data collected from devices with epitaxially grown T centres are shown in black for a single emitter and in orange for an ensemble of emitters.
    Data collected from an ensemble of implanted T centres are shown in purple for comparison. 
    Error bars represent the 1 $\sigma$ uncertainties on the Lorentzian widths obtained from the fits. 
    Markers are measured data, and solid lines represent fits to the data.
    b) Dips in the photoluminescence spectra for the hole-burning measurements used to characterise the instantaneous homogeneous linewidth for total excitation laser power of approximately 20~nW (left panel) and $>$400~nW (right panel).
    Exact values for the input laser power for each dataset are reported next to it. 
    A vertical offset has been added between the plots for ease of visualisation.
    }
    \label{fig:fig3}
\end{figure*}

The SOI wafers with epitaxially grown T centres are diced into individual chips and processed through standard electron-beam lithography techniques to form nanophotonic components to interface with the quantum emitters, as depicted in Fig.~\ref{fig:fig1}f.
We use nanobeam waveguide devices (see Fig.~\ref{fig:fig1}g), in which different excitation lasers can be coupled into the nanophotonic waveguide via an input grating coupler.
The coupled laser light is transmitted through the waveguide and performs resonant excitation of the T centres coupled to the waveguide mode. 
The photons emitted in the waveguide by the excited T centres then propagate to the output grating coupler, which couples them out of the chip for collection.
%
%
Devices are placed inside a dilution refrigerator with a typical operating temperature of 100~mK.
They are optically accessed through windows connecting the sample stage inside the cryostat to a room-temperature optical setup (see Appendix~\ref{sec:supp:experiment} for more details on the experimental apparatus). 
Operating the devices at $100$ mK enables the characterisation of the spectral diffusion and homogeneous broadening properties of the T centres in regimes where thermal noise is highly suppressed~\cite{Zhang2025LaserInduced}. 
In Fig.~\ref{fig:fig2}a, we show the resonant photoluminescence excitation (PLE) spectra for ensembles of T centres. The phonon side-band (PSB) is collected for all measurements unless otherwise specified.
The spectra exhibit inhomogeneous broadening, where the emission lines of individual T centres shift due to strain inhomogeneity in the material.
For epitaxially grown T centres, the inhomogeneous bandwidth is measured as $\Delta^{\text{(epi)}}_\text{inhom} = 30.3(5)$~GHz (orange line). 
This value is comparable to the inhomogeneous broadening of $\Delta^{\text{(impl)}}_\text{inhom} = 24.6(4)$~GHz observed in a second sample, where T centres were produced via standard implantation (purple line, see Appendix~\ref{sec:supp:implanted} for details). 
The similarity suggests that stress inhomogeneities in the two samples may stem from the underlying oxide layer.
%


%
In Fig.~\ref{fig:fig2}b, we show the resonant PLE signal obtained from a single nanophotonic waveguide on the epitaxially grown sample.
The density of T centres is sufficiently low to spectrally isolate a single T centre in the waveguide, whose resonant emission spectrum is highlighted in dark colour in Fig.~\ref{fig:fig2}b.
The rest of the PLE spectrum (orange sections in Fig.~\ref{fig:fig2}b) represents contributions from a small ensemble of T centres present in the waveguide, which are spectrally separated from the single emitter we operate.  
A second-order autocorrelation measurement of the emission from the individual epitaxial T centre, exhibiting clear anticorrelation at zero time delay, is shown in Fig.~\ref{fig:fig2}c.
A second-order correlation value of $g^{(2)}(0) = 0.26(3)$ is obtained, well below 0.5, indicating operation of a single T centre at the single photon level. 
We also report in the right panel of Fig.~\ref{fig:fig2}c the values $g^{(2)}(|\Delta t|)$ of all the side-peaks of the auto-correlation measurement for delay times $|\Delta t|$ up to 400~$\mu$s.  
These measurements show an exponential decay from an initial value of the auto-correlation signal of approximately 1.4 at $|\Delta t|\simeq0$ to an uncorrelated $g^{(2)}(|\Delta t|\rightarrow \infty)$ of 1.0 at long delays, with a decay time of $34(5)~\mu\text{s}$.
The decay is due to spectral diffusion of the emitter, which decreases the correlation between having the emitter resonant with the excitation laser at time $t$ and at time $t+\Delta t$. 
The observed decay time is longer but comparable to measurements from implanted devices in SOI nanophotonic cavities reported in previous studies, where a correlation decay time of approximately $20~\mu$s was observed in second-order correlation measurements~\cite{Bowness2025LaserInduced}.
%



\section{Characterisation of homogeneous linewidths}
\label{sec:homogeneous}

Noise processes faster than the photon emission lifetime affect the instantaneous homogeneous linewidth of a quantum emitter. 
These can be characterised through hole-burning measurements ~\cite{deabreu2023waveguide,  Zhang2025LaserInduced, Bowness2025LaserInduced, siegman1986lasers}. 
In these measurements, pulses from two separate lasers are injected simultaneously into the chip to excite the T centres (as depicted in Fig. \ref{fig:fig1}g). 
One of the two lasers is used as a pump, and its frequency is kept resonant in the centre of the inhomogeneous line of the T centre ensemble.
The second laser is used as a probe, and its frequency is scanned across the T centre inhomogeneous line. 
 Pulses from the two separate lasers are set to arrive simultaneously on the chip.
When the emitter is resonant with the pump laser, the excitation from the probe laser at the same frequency is suppressed by saturation. 
When the emitter resonance frequency spectrally wanders away, the emission is not saturated, and the suppression is absent.
This behaviour gives rise to a hole in the emission spectrum when the two lasers are closer than the average homogeneous linewidth of the emitters. The width of the hole thus provides information about the average homogeneous linewidth at a given total excitation power. 
In general, for powers much smaller than saturation power, the homogeneous linewidth is estimated as half of the hole width ~\cite{deabreu2023waveguide,  Zhang2025LaserInduced, Bowness2025LaserInduced, siegman1986lasers}. 
Observed data for the homogeneous linewidths measured through hole-burning are shown in Fig.~\ref{fig:fig3}a.
We report measured values, collected in the same experimental conditions, for the epitaxially grown devices with a single T centre (black marker) and an ensemble of T centres (orange markers), as well as implanted devices with an ensemble of T centres (purple markers). 
Examples of hole-burning dips used to estimate the homogeneous linewidths are also shown in Fig.~\ref{fig:fig3}b for excitation laser powers (measured off-chip, see Appendix~\ref{sec:supp:experiment}) of approximately 20~nW (left panel) and for higher powers $>$400~nW (right panel). 
A significant reduction of the homogeneous linewidth is observed for the epitaxially grown T centres compared to the standard implanted samples measured in the same experiment.
For example, a homogeneous linewidth of $\Delta^{\text{(epi)}}_\text{hom} = 30(10)$~MHz is measured for the epitaxial single emitter at an excitation power of 24~nW, corresponding to a hole width of 60(20)~MHz.
This value is an order of magnitude smaller than the homogeneous linewidth measured on the implanted sample of $\Delta^{\text{(impl)}}_\text{hom} = 310(80)$~MHz at a smaller excitation power of 19~nW. 
No substantial difference is observed in the spectral hole widths measured on the single T centre and the ensemble of T centres in the epitaxially grown sample for a given excitation power. 
%



\section{Discussion}

We have shown that single T centres in silicon-on-insulator can be generated through epitaxial growth. 
The observed reduction in the homogeneous linewidth suggests that a substantial portion of the dominant decoherence processes that affect homogeneous linewidths in implanted T centres is mitigated by epitaxial growth.
However, similar substantial reductions have not been observed for the spectral diffusion, which in previous studies has also been associated with laser-induced charge recombination in unwanted defects~\cite{Zhang2025LaserInduced}. 
This could indicate that defects giving rise to spectral diffusion are mostly surface-based, such as dangling bonds at oxide interfaces, which are not mitigated by epitaxial growth. 
The observed suppression of homogeneous linewidths can, instead, be explained by a removal of bulk defects, whose charge-related processes are typically faster and thus of higher relevance to homogeneous broadening.

The epitaxial generation of single T centres in silicon-on-insulator opens new engineering routes for quantum technologies for colour-centre-based quantum technologies in a CMOS-compatible platform.
%
These include heterostructure design through the incorporation of doped layers during epitaxy, enabling vertical p-i-n structures for efficient Stark tuning and improved optical performance~\cite{dobinson2026spectral, day2024electrical, somaschi2016near, uppu2020scalable, tomm2021bright}.
The next stage, optimising device design parameters, MBE growth and annealing parameters, and using $^{28}$Si and Deuterium as base materials~\cite{ChartrandSilicon28, bergeron2020silicon, deabreu2023waveguide, Kazemi2026}, will provide additional opportunities to advance devices for highly coherent spin-photon interfaces in epitaxial silicon-on-insulator.

\section{Acknowledgements}

This work is supported by the Novo Nordisk Foundation, Grant number NNF22SA0081175, NNF Quantum Computing Programme and Quantum Foundry Copenhagen. The authors further acknowledge funding support from the Research Council of Norway (research grants No. 349807 and 354831). E.B. acknowledges the PNRR MUR project PE0000023-NQSTI. H.L. acknowledges support from the Swedish Research Council (2023-06502). S.P. acknowledges funding from VILLUM FONDEN (Grant No. VIL60743 and VIL78724), the European Research Council (ERC StG ASPEQT, No. 101221875), and Danmarks Innovationsfond research grant No. 4356-00009B (HyperTenQ). 





\bibliography{biblio}


\newpage 
\onecolumngrid
\clearpage

\appendix

\setcounter{secnumdepth}{2}

\setcounter{figure}{0}
\setcounter{table}{0}

\renewcommand{\thefigure}{\Alph{section}\arabic{figure}}
\renewcommand{\thetable}{\Alph{section}\arabic{table}}


\section{Annealing tests}
\label{sec:supp:anneal}
\begin{figure}[h]
    \centering
    \includegraphics[width=0.5\linewidth]{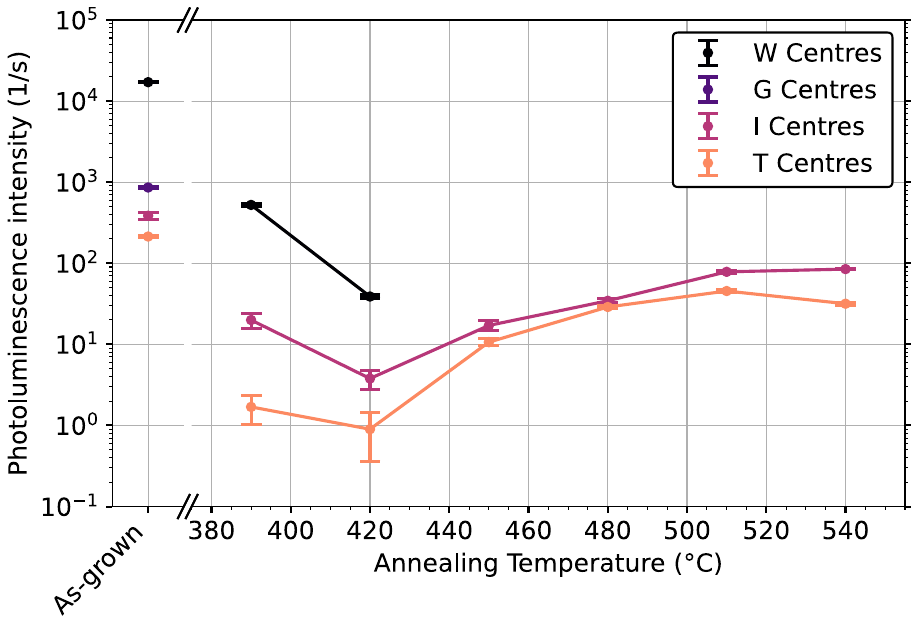}
    \caption{
    \textbf{Photoluminescence signal as a function of annealing temperature}.
    The signal intensity is extracted as the area of a Gaussian fitted to a second-order polynomial background over a window around the PL peak of each colour-centre emission line. Error bars denote $\pm 1\sigma$ statistical uncertainty.
    }
    \label{fig:AnnealSweep}
\end{figure}

To investigate the robustness of the grown emitters to high-temperature processing, we anneal a series of samples from an epitaxially grown wafer with carbon defects at temperatures ranging from \SI{390}{\celsius} to \SI{540}{\celsius}.
For all samples, we use an annealing time of 3 minutes in a nitrogen atmosphere ~\cite{snedker2026colour}. 
PL measurements are then taken from the samples under 10 mW illumination with a 405 nm laser using the setup previously described in Ref. ~\cite{snedker2026colour}. 
The results are shown in Figure~\ref{fig:AnnealSweep}, and they are similar to results from implanted samples, only with a smaller family of present colour centres. 
Similar testing indicates that the stability to fabrication processes is also similar to that of implanted samples ~\cite{snedker2026colour}.

\section{Experimental setup}
\label{sec:supp:experiment}

\begin{figure}[]
    \centering
    \includegraphics[width=1.0\linewidth]{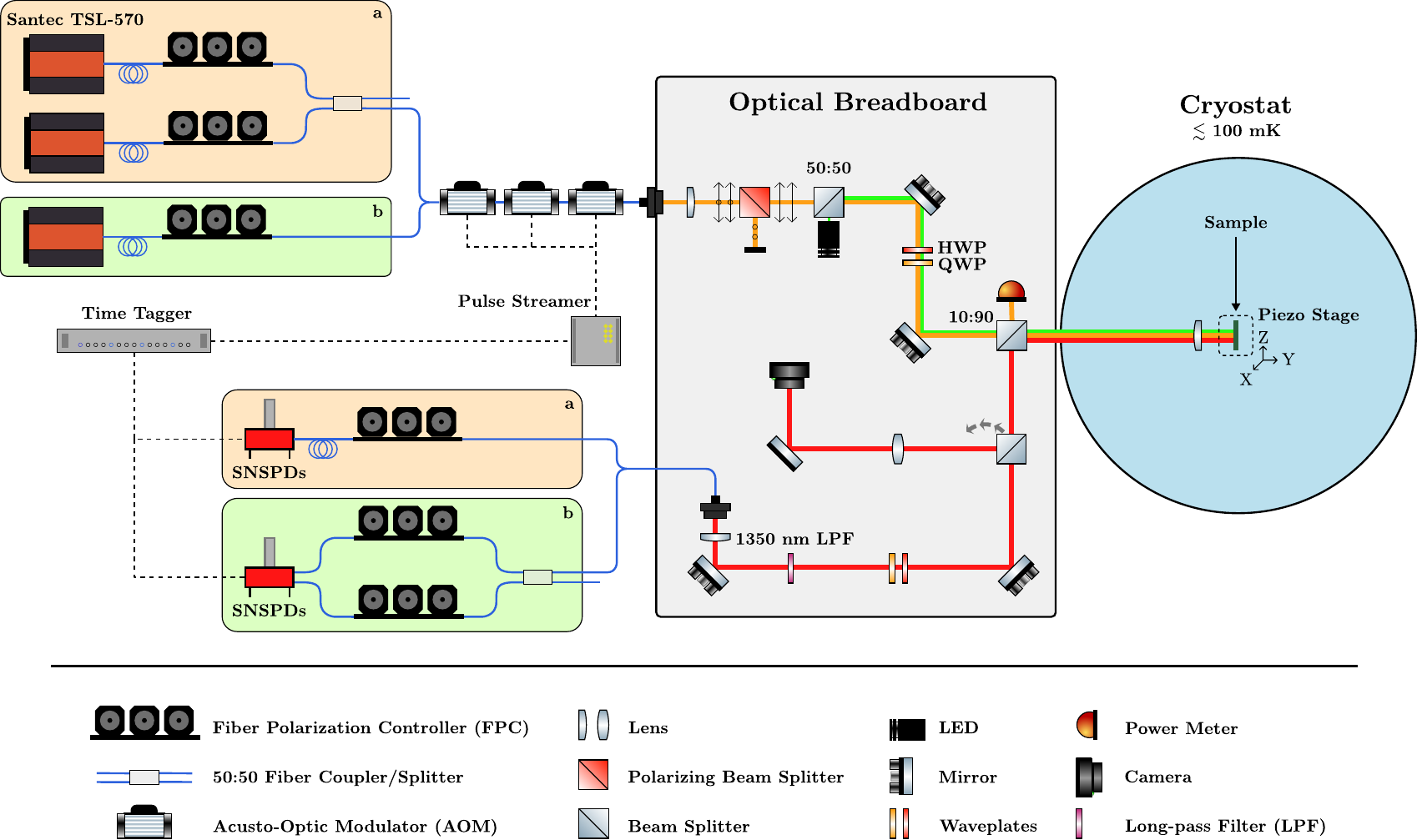}
    \caption{Schematic of the experimental setup. Narrow-width tunable lasers are pulsed, collimated and aligned onto the sample. The emitted light is collected, collimated and directed to superconducting nanowire single-photon detectors (SNSPDs). (a) Resonant excitation and detection scheme used for spectral hole-burning measurements, in which two tunable lasers are combined in fibre prior to pulsing. (b) Resonant excitation and detection scheme used for second-order correlation measurements, where two SNSPDs are arranged in a Hanbury Brown–Twiss (HBT) configuration.  }
    \label{fig:exp_setup_scheme}
\end{figure}

A schematic of the experimental setup is shown in Fig. \ref{fig:exp_setup_scheme}.
The setup used for spectral hole-burning measurements (Fig.~\ref{fig:exp_setup_scheme}a) consists of two tunable continuous-wave lasers (Santec TSL-570), whose polarisations are independently controlled using fibre polarisation controllers (FPCs). One laser (pump) was held at a fixed wavelength, while the wavelength of the second laser (probe) was tuned.
The laser outputs are combined using a 50:50 fibre coupler and pulsed using a sequence of three fast-switching fibre-coupled acousto-optic modulators (AOMs; AeroDiode 1310-AOM-2, rise time $<10$ ns). Pulse timing is controlled by a Swabian Instruments PulseStreamer 8/2. A repetition rate of 125-350 kHz with an optical pulse duration of 200-800 ns was used. 

The pulsed excitation light is delivered to the optical breadboard through a polarisation-maintaining (PM) fibre and subsequently coupled into free space. The beam is first collimated by a lens and then passes through a polarising beam splitter to define a linear polarisation state. A half-wave plate (HWP) and a quarter-wave plate (QWP) are used to optimise the input polarisation for the grating couplers. The beam then passes through a 90:10 beam splitter and is focused onto the sample. The beam splitter reflects most of the excitation light away from the cryostat and towards a power meter. This configuration enables the collection of approximately 90 \% of the light emitted by the sample. Since the available excitation power is not a limiting factor in these measurements, the loss in excitation is outweighed by the improved collection efficiency. 

The sample is maintained at temperatures of $T \lesssim 100$ mK in a BlueFors LD400 dilution refrigerator. The chip is mounted on a three-axis piezoelectric nanopositioner (Attocube ANPx101 and Attocube ANPz102). The objective lens inside the cryostat also collimates the emission from the sample, which is then guided out and coupled into a PM collection fibre. A HWP and a QWP are used to align the linearly polarised emission from the output grating coupler with one of the principal axes of the PM fibre. A long-pass filter (LPF; Thorlabs FELH1350) suppresses the resonant excitation laser while transmitting the phonon sideband (PSB) emission.

A pair of motorised mirrors in both the excitation and collection paths enables fine alignment of the optical beams with the grating couplers. This allows the excitation and collection gratings (coupling efficiencies of approximately $10 \%$) to be optimised around 1326 nm and 1380 nm, respectively, enabling resonant excitation of T centres and efficient collection of PSB emission. An LED illumination source and an imaging camera are also incorporated into the setup, allowing visual inspection of the chip. A flip-mounted beam splitter can be removed from the optical path to disable imaging during measurements. 

The collected emission is routed to superconducting nanowire single-photon detectors (SNSPDs; Single Quantum Eos R12) housed in a separate cryostat operating at a base temperature of $T = 2.2$ K. An FPC is used to optimise coupling into the detectors and maximise detection efficiency. Detection events are recorded using a Swabian Instruments Time Tagger X.

Second-order correlation function measurements (Fig.~\ref{fig:exp_setup_scheme}b) were performed using a single excitation laser and two detectors arranged in a Hanbury Brown-Twiss configuration via a 50:50 fibre splitter. Pulsed resonant excitation was used with a repetition rate of 222 kHz, a pulse duration of 150 ns, and a collection window of 4.2 $\mu$s. The detected emission was time-gated to capture the luminescence decay.

\section{Implanted samples}
\label{sec:supp:implanted}
The implanted sample used as a reference was fabricated following the process described in~\cite{snedker2026colour}.
The sample consists of commercial silicon-on-insulator from Shin-Etsu, with 220~nm of Si on 3~\textmu m SiO$_2$ on 725~\textmu m Si, with a resistivity of $\geq 3000$~$\Omega$~cm.
Ion Beam Services implanted carbon and protons at 38 and 9 keV, respectively, with a stoichiometric ratio of 1:1 and a fluence of $7\times 10^{13}$cm$^{-2}$, annealing at \SI{1000}{\celsius} for 20~s in an Argon atmosphere between the implantations.
Devices are then defined with a single round of e-beam lithography, using a CSAR (Allresist Ar-P 6200) resist mask and an anisotropic dry etch, after which they undergo a final round of annealing at \SI{525}{\celsius} for 3~min in a nitrogen atmosphere.
PL measurements show a 50 times larger intensity as measured by the area under a fitted Gaussian for this implantation as opposed to the grown samples.

\section{Autocorrelation measurement}
The background level for the autocorrelation measurement is estimated using Monte Carlo sampling methods, sampling from the measured temporal detection profile of the individual detectors. The time difference between detection events in the two detectors is sampled to estimate the background. This method includes both coincidences between signal and detector dark counts. The experiment repetition rate is pulsed every 4$\mu$s, which also results in a slight overlap between neighbouring pulses, given the lifetime of ~$0.8\mu$s.

\end{document}